# Einstein's Patents and Inventions

Asis Kumar Chaudhuri
Variable Energy Cyclotron Centre
1-AF Bidhan Nagar, Kolkata-700 064

**Abstract:** Times magazine selected Albert Einstein, the German born Jewish Scientist as the person of the 20th century. Undoubtedly, 20th century was the age of science and Einstein's contributions in unravelling mysteries of nature was unparalleled. However, few are aware that Einstein was also a great inventor. He and his collaborators had patented a wide variety of inventions in several counties. After a brief description of Einstein's life, his collaborators, his inventions and patents will be discussed.

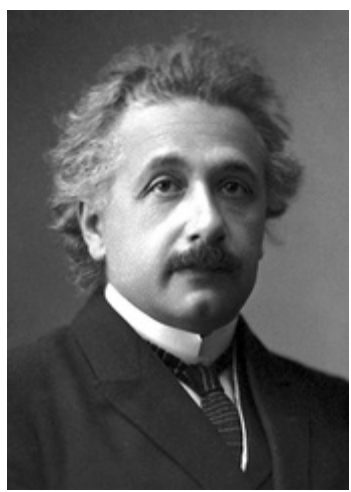

*Figure 1. Einstein in 1921 portrait after receiving Nobel Prize*

## 1. Introduction

Towards the end of the last century, Times Magazine asked some of the World's leading personalities to pick their choice for the person of the century. The magazine compiled a list 100 most influential people of 20th century and the German born scientist Albert Einstein topped the list. Einstein's choice as the person of the century didn't invoke any resentment, it was generally agreed that 20th century is the age of Science and undoubtedly, Einstein's contribution to Science, to the understanding of the intricate laws of nature was unparalleled. He greatly influenced modern science; altered our views on space-time, matter and energy, gave new interpretation to gravity etc. The enormous popularity he enjoyed during his lifetime and even now, is rare for any individual; religious leader, politician, film star. Even a child knows his name, not to speak of adults.

However, while Einstein is known as a great theoretical physicist, few possibly knew that he had more than 50 patents in his names and in several counties. How did a great theoretical physicist get interested in patenting inventions? Here, following a brief description of Einstein's life, we will discuss his inventions and patents.

## 2. Einstein's life in brief

Albert Einstein was born in a Jewish family on March 14, 1879 in a small German city Ulm. His mother, Pauline Koch was a quiet intelligent woman and his father was Hermann Einstein, an engineer and small time manufacturer. At his birth, his mother was worried about Einstein's rather unusual shape of head and the worry increased when Einstein took rather long time to speak. However, he grew up to be a fine gentleman (see the portrait in Figure 1). Pauline had a taste for fine arts. She was also a fine piano player. Throughout his life, Einstein loved music and he inherited his love for music from his mother. Pauline also inspired Einstein to play violin, and he became a fine player too. Shortly after Einstein's birth, Hermann moved to Munich and with his brother Jacob, started a small gas and water installation business. There, on 18 November 1881, Einstein's only sister Maza was born. Throughout his life Einstein was very close to his sister. Hermann and Jacob's gas and water installation business was a success, but Jacob was ambitious and at his insistence, a few years later, they started an electrochemical factory, "Elektrotechnische Fabrik J. Einstein and Co."  The factory produced dynamos, arc lamps, electrical measuring equipment etc. for electric power stations and lighting system. Initially, the factory was a success. In 1885, the company won a contract to illuminate Munich city during the Oktoberfest.[1] However, with the advent of alternating current, company's fortune started to dwindle.  Their factory used to make instruments for direct current. Germany started to convert its electrical installations from direct to alternating current and their instruments had no buyer. The brothers did not have the enough capital to convert their factory to make instruments for alternating current. In 1894, the Munich factory was relocated to Pavia, Italy. Einstein's family also moved to Italy, but to continue his education, Einstein remained in Munich. Pavia factory was also a failure. Though Hermann Einstein started another company and which was a moderate success, he never recovered from the shock of earlier failures and died in 1902, at the age of 55.

     Einstein received his primary education at a Catholic school and later at Luitpold-Gymnasium[2] in Munich. However, he did not like the rigid atmosphere at the gymnasium. When his parents moved to Italy, he was miserable and lonely. Without consulting his father he left the school in 1894, without a degree.  Later, he continued his study in a Swiss gymnasium at Aarau and obtained his school leaving certificate (matura) in 1896.  There was a misconception that as a student Einstein was poor.  It was wrong. For record purpose, I am listing his final grading in matura (maximum=6): German 5. Italian 5, geography 4, algebra 6, geometry 6, descriptive geometry 6, physics 6, chemistry 6, natural history 5, drawing (art) 4, drawing (technical) 4. Having passed matura, he enrolled in the four-year mathematics and physics teaching diploma program at the ETH-Zürich Polytechnic (*Eidgenössische Technische Hochschule Zürich*), a science, technology, engineering and mathematics University at Zurich. His fellow students were:  Marcel Grossman, Louis Kollros, Jacob Ehrat and Mileva Maric. Einstein enjoyed his Zurich days.  He developed a lifelong friendship with Grossman and his family. He courted his future wife Mileva, the only girl in his class. He also acquired his lifelong friend Michele Angelo Besso[3]. The ETH program was completed in August 1900. Einstein and his fellow students, with the exception Mileva Maric[4], passed the examination. His ETH-Zurich scores were (out of maximum 6):  5 each for theoretical physics, experimental physics and astronomy; 5.5 for the theory of functions; 4.5 for an essay on heat conductivity.

---

[1] *Oktoberfest also known as the beer festival, is the world's largest beer festival   held annually in Munich. It is a 16-18-day festival running from mid or late September to the first weekend in October, with more than 6 million people from around the world attending the event every year.*
[2] *Gymnasiums are kind of secondary schools, with strong emphasis on academic learning, preparing students for University education*
[3] *Michele Besso (25 May 1873–15 March 1955) was an Italian engineer and most close friend of Einstein during his years at ETH Zurich, and then at the patent office in Bern, where Einstein helped him to get a job. Einstein used to discuss his ideas with him and called him "the best sounding board in Europe" for scientific ideas. In his first paper on relativity, which did not have any references, he acknowledged Besso, "In conclusion I wish to say that in working at the problem here dealt with I have had the loyal assistance of my friend and colleague M. Besso, and that I am indebted to him for several valuable suggestions."*
[4] *In her second attempt also Mileva failed the examination.*

On leaving Germany in 1894, Einstein also renounced his German nationality, otherwise, on attaining 16 years of age, he had to return back to Germany. In Germany military service was compulsory and all German nationals on attaining 16 years of age, had to report for conscription. From his childhood, Einstein had aversion for the military service. It started when as a child he observed a military march. He was revolted seeing all the soldiers mechanically marching to the tune of band. Later, he expressed his distaste for military service,

*"He who joy fully marches to music rank and file has already earned my contempt. He has been given a large brain by mistake, since for him the spinal cord would surely suffice."*

From 1896 to 1901 Einstein was a stateless person when on payment of 600 Swiss franc, he became a Swiss national. It may be remembered that throughout his student days, his family was in financial trouble and Einstein had to live on an allowance of 100 Swiss franc per month of which he saved twenty to pay for his Swiss naturalization papers. Einstein's allowance was modest not meager. In 1895-1900, typical cost for living at Zurich was Swiss franc 60-80 per month.

After his graduation Einstein had trouble finding a suitable job. To his great disappointment, three of his fellow students (Marcel Grossman, Louis Kollros, Jacob Ehrat), were offered teaching positions at ETH. In particular, he accused Professor Heinrich Friedrich Weber[5] for holding out an assistantship. In 1902 through the influence of Grossman's father he secured a job in the Swiss patent office. On June 16, 1902 he was appointed technical expert third class at the patent office at Bern with an annual salary of SF3500. Four years later, he was promoted to technical expert second class with salary SF4500. Patent office director Friedrich Haller considered him to be one of the most esteemed experts at the office. From his ETH days, Einstein was courting his fellow student Mileva. Now, with a permanent job, he wanted to marry her. However, initially, Einstein's family was opposed to the union. They were Jew while Mileva came from an orthodox Greek family. Eventually, Einstein could prevail upon his parents and on January 6, 1903 he married Mileva. The couple had three siblings: a daughter, Lieserl (died early at the age of one year) and two sons, Hans and Eduard Einstein. However, the marriage was not a success. Mileva was frustrated as her own career as a physicist did not develop and in 1919, they agreed to divorce. In 1919 Einstein was famous but he was yet to receive the Nobel Prize. He was so certain to get it that as part of divorce compensation, he signed over the award money to his wife, Mileva. In 1922, when Einstein was awarded the prize, the prize money 121,572:54 Swedish kronor, equivalent of more than twelve years' income for Albert Einstein, was transferred to Mileva's account. Five months after the divorce, Einstein married his cousin Elsa. Elsa was also a divorcee with two daughters and unlike Mileva, was not a brilliant woman. She took good care of Einstein and little by little she became an important part of Einstein's life.

Until 1905, Einstein's life was rather featureless. He diligently worked at the patent office, played violin, discussed physics with his friends, write few not so interesting papers. Then in 1905, he took the academic world by surprise. In the annals of physics, the year 1905 is known as "annus mirabilis" or the year of miracle. Indeed, a miracle happened. Albert Einstein, barely 26 years old, sitting in an obscure Swiss patent office, wrote four papers, each of which produced some sort of revolution in Physics. The papers were published in a single issue (issue no.17) of the reputed German journal of Physics, Annalen der Physik (publishing since 1799**).** Below a brief description of the papers are given.

---

[5]*Heinrich Weber was a German physicist who made important contributions to measurement of specific heat. For reasons unknown, Prof. Weber took a dislike of Einstein and once commented," You are a smart boy, Einstein, a very smart boy. But you have one great fault: you do not let yourself be told anything." When in 1912 Weber died, Einstein wrote to a friend, in a way quite uncommon for his, 'Weber's death is good for ETH.'*

*1. On a heuristic viewpoint concerning the production and transformation of Light. Annalen der Physik 17 (1905)132-148.*

In this paper, motivated by Max Plank's black body radiation law, Einstein proposed the idea of energy quanta; luminous energy can be absorbed or emitted only in discrete amounts. He then explained the photoelectric effect[6]. We may add that eventually, this paper led Einstein to the coveted Nobel Prize in Physics.

*2. On the movement of small particles suspended in stationary liquids required by the molecular theory of heat, Annalen der Physik 17 (1905) 549-560.*

This is Einstein's first paper on Brownian motion. Einstein derived the relation between diffusion coefficient and viscosity, re-derived diffusion equation and expressed Avogadro's number in experimentally determined quantities.

*3. On the electrodynamics of moving bodies, Annalen der Physik 17 (1905) 891-921.*

This is the first paper on special relativity. It drastically altered the century old man's idea about space and time. In Newtonian mechanics they have separate identities. In Einstein's relativity, space and time are not separate entities rather one entity called space-time continuum. Continuum because in our experience there is no void in space or time. Identification of space-time as an entity required that bodies moving with velocity of light or near need a different mechanics, relativistic mechanics rather than the Newtonian mechanics. Intermingling of space and time produces few surprises, e.g. a moving clock tick slowly (time dilation), a moving rod contracts (length contraction), strange laws of velocity addition etc.

*4. Does the inertia of a body depend upon its energy content" Annalen der Physik 17 (1905) 639-641.*

This is the second paper on relativity. He derived the mass energy relation: $E=mc^2$, the most well-known mathematical equation in the World.

In addition to the four papers mentioned above, in 1905, Einstein, under the supervision of Alfred Kleiner, also submitted his thesis for PhD degree; *"A new determination of molecular dimensions."* The thesis was 24 page only. Einstein recounted that his supervisor, Kleiner returned the thesis saying it was too short. Einstein added a single sentence and it was accepted without further comments.

The full import of Einstein's theory took time to percolate even in the academic world. Only few realized their true values, among them was Wilhelm Wein[7], editor of Annalen der Physics.

---

*[6]In photoelectric effect, certain metals when shined by light emit electrons. Einstein, using quantum mechanics gave proper explanation of the phenomenon. Indeed, it was the first application of quantum mechanics in a physical process other than radiation.*

*[7]Wilhelm Wein (1864-1928) was a German physicist who did pioneering work on radiation. He discovered the displacement law (the wavelength changes with temperature). He also introduced the concept of ideal or black body (the body which completely absorbs all radiation). He also obtained the black body radiation law known as Wein's law, which later found to be valid only for short wavelength*

Immediately after the 17th issue was published he came to Johann Laub[8] and asked him to discuss Einstein's papers in the next colloquium. Max Plank[9] at Berlin also realized their importance. He was one of the few who understood relativity. Immediately, after the paper was published he gave a colloquium on relativity. The other man was the Polish professor Witkowski[10], who after reading Einstein's paper proclaimed to his colleagues, "A new Copernicus is born! Read Einstein's paper." Even though a few recognized his ability, he remained largely unknown to the academic world. For example, even in 1907 Max Born[11] apparently was not aware of Einstein's papers. In 1907, Einstein applied for a Privatdozentship at the University of Bern. Privatdozentship carried no salary, but only right to teach. However, it was denied due to a technical flaw in the application and only after Einstein corrected the technical flaw, he was granted Privatdozentship and formally, Einstein became a member of the academic world. In the meantime, he even contemplated for applying for a teacher's position in a school. Gradually, his reputation as a mathematical physicist grew and he was offered his first faculty position, associate professor of the theoretical physics at the University of Zurich. Einstein left his job at the Patent office and joined the University of Zurich on October 15, 1909. Thereafter, he continued to rise in ladder. In 1911, he moved to Prague University as a full professor, a year later, he was appointed as full professor at ETH, Zurich, his alma-mater. In 1914, he was appointed Director of the Kaiser Wilhelm Institute for Physics (1914–1932) and a professor at the Humboldt University of Berlin, with a special clause in his contract that freed him from teaching obligations. In the meantime, he was working for a theory of gravity. The work started in 1907 and after eight long years, in 1915, he could finalize his theory of gravity. He christened it as the General Theory of Relativity. In general relativity, Einstein gave a geometric picture of gravity: in presence of mass, space-time is curved, and gravity is nothing but curvature of the space-time. In 1921, he received Nobel Prize in Physics for the explanation of the photoelectric effect. In 1933, when Hitler assumed power and started persecuting Jews, Einstein left Germany and joined Princeton University in USA, where he continued until his death on April 18, 1955. In his later years, he tried to unify all the fundamental forces but could not succeed.

One of the consequences of general relativity was that light will bend around a massive object. Indeed, bending of light was also a prediction of Newtonian gravity, but in general relativity the bending is approximately twice the bending in Newtonian gravity. In May 29, 1919, during the total solar eclipse, Sir Aurther Eddington[12] and his collaboration measured the bending of light due to Sun,

---

*radiation. In 1911, he was awarded Nobel Prize in Physics for his discoveries regarding the laws governing the radiation of heat.*

[8] *Jacob Johan Laub (1884-1962) was an Austria-Hungary physicist. He was educated in University of Vienna, University of Krakow, University of Goettingen. In 1905, he was working with Wilhelm Wein on Cathoode rays. Once he was acquainted with Einstein's theory, he started investigation of special relativity and wrote in 1907 an important work on the optics of moving bodies.*

[9]*Max Plank (1858-1947) was a German theoretical physicist who pioneered quantum mechanics. He is best known for discovering energy quanta and was awarded the 1918 Physics Nobel Prize in recognition of the services he rendered to the advancement of Physics by his discovery of energy quanta.*

[10]*August Witkowski (1854-1913) was a Polish physicist at the University of Krakow, Poland. He was mainly experimental physicist and studied physical properties and laws concerning gases, especially in low temperature. In later years, he worked on relativity. His three volume Handbook of Physics were basis of academic education for several generation of Polish scientists.*

[11] *Max Born (1882-1970) was a German physicist and mathematician who was instrumental in the development of quantum mechanics. In 1954, he was awarded Nobel Prize in Physics for his "fundamental research in Quantum Mechanics, especially in the statistical interpretation of the wave function."*

[12] *Sir Arthur Eddington (1882-1944) was an English astronomer, physicist, and mathematician. He was one of the admirer of Einstein and played an important role in popularising Einstein's works to the*

simultaneously in the cities of Sobral, Ceará, Brazil and in São Tomé and Príncipe on the west coast of Africa. The observed bending beautifully agreed with Einstein's prediction.

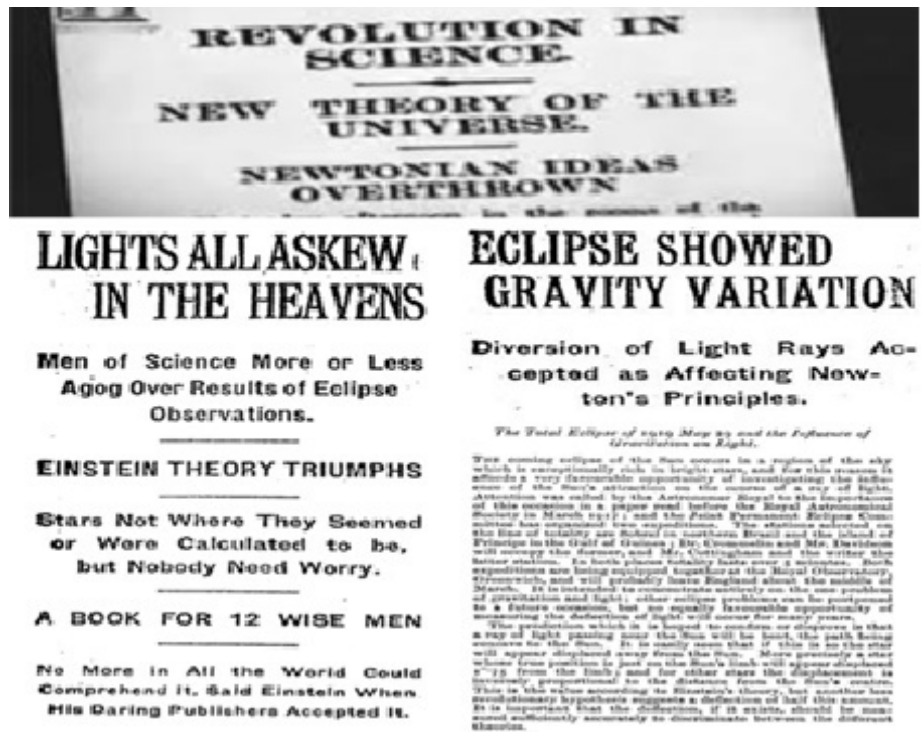

*Figure 2. Newspapers reporting bending of light during 1919 solar eclipse.*

Now, few words are necessary here to understand the social condition of Europe in May, 1919. World War I, which devastated Europe for four long years formally concluded in 11 November 1918. It was one of the deadliest conflicts in the World's history, over 9 million soldiers and seven million civilians died in the war. More than 60 million Europeans suffered untold misery. In the beginning of 1919, Europe was a depressingly gloomy place. It needed some excuse to cheer up. Einstein's prediction that light will bend in vicinity of massive object and its verification by Eddington provided the much needed excuse. The event was reported in all major newspapers (see some of the clipping in Figure 2) and overnight Einstein became a household name. Relativity caught the imagination of the general populace and many stories and limericks were written. One such limerick was published in Punch in 1922:

*Relativity.*

*There was a young lady named Bright*
*Whose speed was far faster than light;*
*She set out one day*
*In a relative way*
*And returned on the previous night.*

---

*English speaking world. He also played an important role in the life of the Indian Nobel Laureate Subrahmanyan Chandrashekhar's life by ridiculing his concept of black hole.*

## 3. Einstein's Inventions and Patents

*Table 1: Patents of Jacob Einstein (Albert Einstein's uncle).*

| Date | Patent number | Collaborators | Description |
|---|---|---|---|
| 30/08/1890 | CH2131 | Sebastian Kornprobst | New electrical measuring and registering apparatus |
| 31/12/1886 | DE41824 | J. A. Essberger | Improvements in electric arc lamps |
| 30/11/1889 | DE53207 | - | Automatic circuit breaker for electric arc lamps |
| 26/02/1890 | DE53546 | Sebastian Kornprobst | Apparatus for stabilising irregular indicator movement in electric meter displays |
| 21/11/1889 | DE53846 | Sebastian Kornprobst | Improvements to electric measurement apparatus |
| 23/02/1890 | DE60361 | Sebastian Kornprobst | Spring loaded friction wheel |
| 10.10/1893 | DE74429 | | Control of carbonisation in electric arc lamps |

*DE-Germany, CH-Switzerland*

Considering Einstein's upbringing, his interest in inventions and patents was not unusual. Being a manufacturer's son, Einstein grew upon in an environment of machines and instruments. When his father's company obtained the contract to illuminate Munich city during beer festival, he was actively engaged in execution of the contract. In his ETH days Einstein was genuinely interested in experimental works. He wrote to his friend, "*most of the time I worked in the physical laboratory, fascinated by the direct contact with observation.*" Einstein's Uncle Jacob was enterprising and obtained several patents on electrical instruments (see Table 1). He was aware that his father's factory run into trouble due to introduction of new technology or new invention. Thus, Einstein was no alien to inventions and patents. He clearly understood the connection between invention, technology and patent. Moreover, his experience in the patent office, where he evaluated hundreds of patent applications made him an expert in nuances of machines and instruments. However, it must also be emphasized that his main occupation was theoretical physics. The inventions he worked upon were his diversions. In his unproductive times he used to work upon on solving mathematical problems (not related to his ongoing theoretical investigations) or took upon some practical problem. As shown in Table. 2, Einstein was involved in three major inventions; (i) refrigeration system with Leo Szilard, (ii) Sound reproduction system with Rudolf Goldschmidt and (iii) automatic camera with Gustav Bucky. He also obtained a patent for a design of a blouse. It must also be mentioned that none of Einstein's inventions came to consumer market and with the exception of the refrigeration system, they are of historical importance only. Below, his inventions are discussed in detail.

### 3A. Einstein-Szilard Refrigeration system

Among all of Einstein's inventions, possibly the most important was design of a refrigeration system with Leo Szilard. They obtained maximum number of patents on it, possibly due to the zeal of Szliard. Leo Szilard (1898-1964) was a theoretical physicist and a fellow Martians[13]. Leo Szilard was born in a Jewish family in Budapest, Hungary. His father was an engineer. On his father's advice, Szilard, after honourably completing school education (he obtained the Eötvös prize, a national prize for mathematics) enrolled in a technical university for a degree in electrical engineering. World War I intervened his study and he joined the artillery service. After the war he resumed his engineering study in Berlin University. There he found his true interest: Physics. The physics faculty of Berlin University then housed such luminaries like Albert Einstein, Max Planck and Max von Laue. Strongly influenced by those luminaries Szilard abandoned engineering and switched to Physics. He did a very important work on thermodynamics for his Ph.D. degree under von Laue. After his PhD, he joined Kaiser Wilhelm Institute as a Privatdozent and continued there until 1933. In 1933, with rise of Nazism, he left Germany and went to England. In 1938 he obtained a teaching position at the Columbia University, US. Later, he moved to Chicago to work with Enrico Fermi on the first Nuclear Reactor project. Szilard played an important role in the making of atom bomb. Concerned that Hitler's Nazi Germany may make one, he and Eugene Wigner (a fellow Hungarian physicist) impressed upon Einstein to write a letter to US President Roosevelt warning him of such a possibility and suggesting that USA should start its own Nuclear Science programme. The letter was drafted by Szilard and Einstein signed it. With the letter began the Manhattan project which ultimately built the Atom Bomb. Szilard was a gifted scientist. In 1933, long before nuclear fission was discovered, he conceived the idea of nuclear chain reaction to extract energy from nucleus and later, patented the idea with Enrico Fermi. In 1928, he patented the idea of Cyclotron, but before he could proceed further, Ernest Lawrence of Berkeley made a small cyclotron. Near the end of World War II, while Japan was negotiating for surrender, on August 6, 1945, America dropped the first atom bomb on the Japanese town Hiroshima and three days later on August 9, 1945 on Nagasaki. Szilard tried his best to desist America from dropping the bombs but could not. He was dejected by the devastation caused by the bombs and in a way held himself responsible. He shifted his interest from Physics to biological science. There also he excelled, he invented chemostat, bioreactor where specific growth rate of microorganism can be easily controlled; discovered feedback inhibition and was involved in first cloning of human cell. In 1964, he died while sleeping.

---

[13]*A group of Jewish-Hungarian scientists who immigrated to the US in the first half of 20th century came to be known as "The Martians." The group included, Theodore von Karman (outstanding aerodynamic theoretician of the twentieth century), John von Neumann (regarded as the foremost mathematician of his time, pioneered application of operator theory in quantum mechanics, game theory, cellular automata and digital computer), Paul Halmos (outstanding mathematician and mathematical expositor), Eugene Wigner (theoretical physicist and mathematician, received Nobel prize in Physics in 1963 "for his contributions to the theory of the atomic nucleus and the elementary particles, particularly through the discovery and application of fundamental symmetry principles), Edward Teller (theoretical physicist came to be known as father of the hydrogen bomb), George Polya (mathematician, his book, 'Problems and Theorems in Analysis' influenced generation of physicists and mathematicians), Denis Gabor (engineer-physicist, 1971 winner of Nobel prize in Physics for inventing holography), Paul Erdos (most prolific mathematician of 20th century, known for his works in number theory and combinatorics. Authored more than 1500 papers, worked with more than 500 collaborators. Known for his eccentric life-style, didn't have a permanent position and address) among others. The name was jocularly suggested by Szilard. Once, there was a discussion about extra-terrestrial intelligent life. With myriads of stars in the universe, with many of them not unlike our sun, it is expected that earth like planets with intelligent being are abundant in the universe. If that is so, why there is no evidence? With his impish sense of humor, responded, "They are already here among us: they just call themselves Hungarians."*

Table 2: Albert Einstein's patents with his collaborators.

| Collaborator | Date | Patent No. | Description |
|---|---|---|---|
| Leo Szilard | 01/12/1928 | FR647838 | Refrigerating machine with pumping of liquid effected by intermittently increasing the vapour pressure. |
| | 28/11/1929 | FR670428 | Refrigerating machine |
| | 15/11/1928 | GB282428 | Improvements relating to refrigerating apparatus. |
| | 26/06/1930 | GB303065 | Electrodynamic movement of fluid metals particularly for refrigerating machine. |
| | 09/03/1931 | GB344881 | Pump especially for refrigerating machines. |
| | 05/12/1929 | HU102079 | Refrigerator |
| | 11/11/1930 | US102079 | Refrigeration |
| | 26/05/1933 | AT133386 | Condenser for refrigeratot |
| | 16/08/1930 | CH140217 | Refrigerator |
| | 27/07/1933 | DE554959 | Apparatus for movement of fluid metals in refrigerators |
| | 04/07/1933 | DE565614 | Compressor |
| | 30/05/1933 | DE563403 | Refrigerator |
| | 08/04/1933 | DE562300 | Refrigerator |
| | 20/09/1933 | DE562040 | Electromagnetic appliance for generating oscillatory motion |
| | 13/04/1933 | DE561904 | Refrigerator |
| | 16/09/1933 | DE556535 | Pumps especially for refrigerators |
| Gustav Bucky | 27/10/1936 | US2058562 | Light intensity self-adjusting camera |
| Rudolf Goldschmidt | 10/01/1934 | DE590783 | Electromagnetic sound reproduction apparatus |
| | 27/10/1936 | US101756S | Design of a blouse |

FR:France, AT:Austria, CH:Switzerland, DE:Germany, GB:Great Britain, HU:Hungary, US:USA

From 1926 until 1933 Einstein and Szilárd collaborated on ways to improve home refrigeration technology. Basic principle of refrigeration is simple. You pass colder liquid (called refrigerant) around the object to be cooled. The liquid will take away the heat from the object and cool it. For efficient refrigeration commercial refrigerators generally compresses a refrigerant gas (e.g. Freon etc.) to liquefy it. The process heats up the refrigerant. It is then released into a large chamber, where sudden pressure drop evaporate the refrigerant and its temperature drops. The cool refrigerant then exchanges heat with the target chamber. In Einstein's time, sulphur dioxide, methyl chloride, ammonia are generally used as refrigerant and they were toxic. Accidents due to leak of refrigerant was not uncommon. In 1926 a newspaper reported that all members of a Berlin family, including children, died while sleeping as their refrigerator leaked toxic fume. Einstein was deeply affected by the tragedy, and together with Szilard set out to design a safe refrigerator without any moving part by eliminating the mechanical compressor and toxic gases. They built an absorption refrigerators. In absorption refrigerator, a heat source –in that time, a natural gas flame–is used to provide the energy needed to drive the cooling process. In 1922 Swiss scientists, Baltzar von Platen and Carl Munters invented the technology, and Szilard found a way to improve on their design.

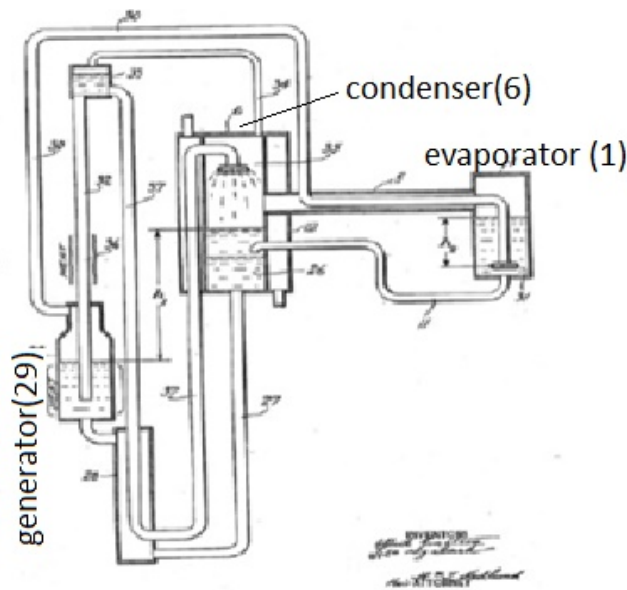

*Figure 3: Schematic diagram of Einstein-Szilard refrigeration system.*

Schematic diagram of Einstein-Szilard refrigerator is shown in **Figure 3**. It is based on the simple physics idea that liquids boil at lower temperatures when the surrounding air pressure is low. For example on top of a mountain water boils at lower temperature than at sea level. The working principle of their refrigerator is explained below:

Butane, the refrigerant, in liquid form, is contained within evaporator (1) and an inert gas, ammonia, is introduced into the evaporator through a conduit. Ammonia lowers the partial pressure of the refrigerant and butane boils taking the energy from the surroundings and lower the temperature. The gaseous mixture of ammonia and butane passes through a conduit into the condenser (6). Through a conduit water is introduced into the condenser. Ammonia is dissolved in water, but butane is not. However, pressure and temperature in the condenser is sufficient to liquefy butane and as the specific gravity of liquid butane is less than that of the ammonia solution, it floats upon the ammonia solution. The liquid butane is then returned to the evaporator (1) through the conduit (11). There, it is again evaporated and the cycle repeated. The ammonia solution flows by gravity from condenser (6) through a conduit and heat exchanger jacket to within generator (29). Here the application of heat causes the ammonia to be expelled as a gas from the solution and this ammonia gas passes through a conduit to the evaporator (1). As before, it reduces the partial pressure of the butane causing its evaporation. The water is again passed to the condenser.

The heart of Einstein-Szilard refrigerator design was an electromagnetic pump. They wanted a built a pump without any moving parts, no gaskets or seals that may fail. Both Einstein and Szilard became intrigued by the problem; how it can be made? Then they thought about human heart. In heart spasmodic contraction of muscles produces pump action. They wanted to duplicate the spasmodic action using electromagnetic waves. The basic principle was simple; when a current carrying conductor is subjected to a magnetic field, a force is produced in a direction perpendicular to the direction of the current and magnetic field. Einstein once outlined the idea as follows: "*an alternating current produces an alternating magnetic field that keeps a fluid potassium-sodium alloy moving. The fluid alloy performs an alternating motion inside a closed housing and acts as the piston of a pump for the refrigerant which is thus mechanically condensed and upon re-evaporation generates cold.*"

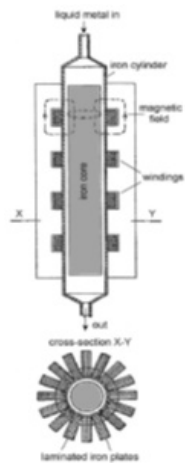

*Figure 4: Schematic diagram of Einstein-Szilard Electromagnetic pump.*

The schematic diagram of their pump is shown in Figure 4. In their pump, liquid metal (mercury) flows through the annular space between an iron core and an iron cylinder under the influence of magnetic fields produced by electric coils surrounding the cylinder and an electric current is induced in it by the magnetic fields. The currents in the adjacent coils are 90 degree out of phase with each other.

Despite filing more than 45 patent applications in six different countries, none of Einstein-Szilard's refrigerators ever became a consumer product. Electrolux Servel Corporation of New York purchased the patent design and Einstein-Szilard obtained a tidy sum $750 (the equivalent of $10,000 today). Efficiency of the refrigerator was rather poor. The great depression of 1930's also hurt many of the manufacturer. Moreover, in 1930, introduction of Freon; non-toxic refrigerant spelled the doom for Einstein-Szilard refrigerator. The German company AEG (Allgemeine Elecktrizitats-Gesellschaft) agreed to develop the pump technology. Szilard and a company engineer Albert Korodi built a prototype pump. But the device was extremely noisy (a contemporary researcher said it "howled like a jackal" ). Another challenge was the choice of liquid metal. Mercury wasn't sufficiently conductive, so the pump used a potassium-sodium alloy instead, which required a special sealed system because it is so chemically reactive. In 1950, the pump found its use in nuclear industry, in high breeder reactors.

## 3B. Sound reproduction system with Rudolf Goldschmidt

Rudolf Goldschmidt was a German Engineer and inventor. He earned his engineering degree in 1898 and PhD in 1906. He spent a decade working in England with major firms such as Crampton, Arc works, Westinghouse etc. On returning back to Germany he joined Darmstadt T H University as a professor. Goldschmidt was a prolific inventor. His first patent was for a bicycle gear while still an engineering student. In 1908 he developed a rotating radio-frequency machine, which was used as an early radio transmitter. The transmitter was used in the first trans-Atlantic radiotelegraphic link between Germany and United States, opened on 19th June, 1914, with an exchange of telegrams between Kaiser Wilhelm II and President Woodrow Wilson.

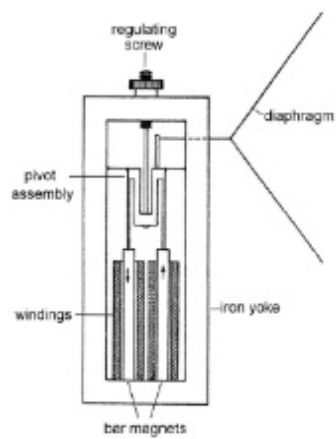

*Figure 5: Einstein-Goldschmidt design of a sound reproduction system.*

In 1922 Goldschmidt approached Einstein for his expert opinion regarding one of his patents. Thereafter, they kept in touch. Even after Einstein migrated to US and Goldschmidt to England, the friendship continued. On January 10, 1934 Goldschmidt and Einstein patented an "Electromagnetic sound reproduction apparatus." The patent has a history. One of the acquaintance of Einstein, Olga Eisner, a distinguished singer, became hard on hearing, a serious drawback for a singer. In 1928, Einstein asked Goldschmidt's assistance in developing a new type of hearing aid for her. At that time he sent his friend one of his poetic creations:

> Ein biszchen Technik dann und wann
> Auch Gruebler amusieren kann
> Drum kuehnlich denk ich schon so weit:
> Wir legen noch ein zu zweit.

In translation,
> A bit technique now and then
> Can also amuse thinkers.
> Therefore, audaciously I'm thinking far ahead:
> One day we'll produce something good together.

Their collaboration resulted a design for sound reproduction system and a patent for the system. The schematic diagram of the device is shown in Figure 5. The final patent was entitled, "Device, especially for sound-reproduction equipment, in which changes of an electric current generate movements of a magnetised body by means of magnetostriction." As the title suggests, they used the physical phenomena of magnetostriction. What is magnetostriction? If a coil is wound round an iron magnet and a current is passed, the rod's magnetism changes, so does the length. The change is small due to rod's rigidity, which counters the magnetostriction. Einstein and Goldschmidt's idea was to diminish the rigidity by keeping the rod under extreme pull to reduce it to a state of labile (where the external pull or pressure is close to its tear or buckling), such that the change in the length of the rod follows the changes in the electric current. When this arrangement is made in a microphone or loud speaker, lengthwise vibration of the rod follows the oscillation of the current in the coil produced by the sound to be transmitted.

However, the original plan of development of a hearing aid did not materialize. They tried to invent an electro-acoustical hearing aid. The basic idea was to convert the acoustical signal into electrical oscillation and transmit the signal by some sort of membrane attached to the skull such that the bone could conduct it to the hearing organ. They worked on the design but both had to flee Germany and their work was disrupted. Later, Goldschmidt tried to revive the idea, but Einstein was not interested. Furthermore, development of electronic hearing aid (where the audio signal is amplified many times) overshadowed their electro-acoustical device.

### 3C. Light Intensity self-adjusting Camera with Gustav Bucky

Einstein, with his long-time friend Gustav Peter Bucky invented a self-adjusting camera few years before Kodak introduced Super Six-20, the generally known as the first automatic camera. It may be mentioned here that Kodak's camera operated on a principle different than that of Einstein-Bucky's camera. Incidentally, this is the only invention where Einstein used a physical phenomenon; the photoelectric effect, which he had investigated theoretically.

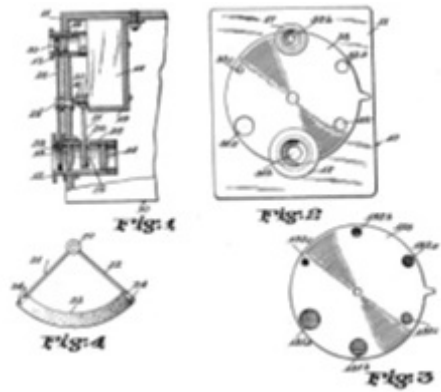

*Figure 6: Schematic design of Einstein-Bucky's automatic camera. 1 shows the side elevation of a portion of a camera as per the invention, partly in section, 2 a front elevation of the camera, 3 a front elevation of a modified part, and 4 an elevation of the screen.*

Gustav Bucky was a German-American radiologist who is famous for discovering Bucky Grid, which enhanced diagnostic ability of X-ray manifold. He was born in a Jewish family in Leipzig. He was interested to study engineering, but at the insistence of his parents, forced to study medicine. From his childhood, he was interested in photography, and as a medical practitioner, he got attracted to radiology. In those days, diagnostic ability of X-rays were limited due to scattered x-rays blurring the film. In 1913, Bucky invented a system of two plates with grids on them. One plate was placed between the X-ray beam and the patient, and the other was placed between the patient and the film. The grids ensured that the secondary particles stayed in columns rather than scattering across the X-ray field.

The Bucky grid reduced the blur of X-ray images, but it caused grid lines to appear on the films. He patented the idea in US and Germany. Worsening political situation forced him to leave Germany in 1933 and settled in US. Bucky became close friend on Einstein in the course of providing treatment to his wife Elsa.

On 11[th] December, 1935, Einstein along with Gustav Peter Bucky (1880-1963) filed a patent application for the self-adjusting camera in US. The patent was granted on October 27, 1936. The patent application was accompanied with diagram (see Figure 6.) and detailed description of the working principle. The purpose of the invention was stated as, "*to provide means for automatically adapting the light impinging the photographic plate or film of the camera to the light intensity of the surroundings and particularly of the object to be photographed.*" In the camera, a photoelectric cell (Weston photronic photoelectric cell) together with a mechanism for turning a shaft was suitably mounted. As the shaft rotates, it moves a screen of varying transparency before the main camera lens system. The light intensity striking the photo-electric unit creates a power which turns the shaft through an angle which is function of the light intensity.

**3D. Einstein's Design of a Blouse:**

It is rather amusing to note that Einstein was interested in cloth design. In 1935, in a letter, Gustav Bucky complained to Einstein that Emil Mayer, an attorney, on behalf of Einstein and Gustav Bucky, has applied for patent for a waterproof garment (without their consent). Possibly, patent application was revoked.

However, records shows that in 1936, Einstein was granted a US patent for a design of a blouse. He applied for the patent on July 2, 1936 and it was granted on October 27, 1936. The design submitted for the patent is shown in *Figure 7*. The design was characterized by the side openings which also serve as arm holes; a central back panel extends from the yoke to the waistband as indicated in the figure.

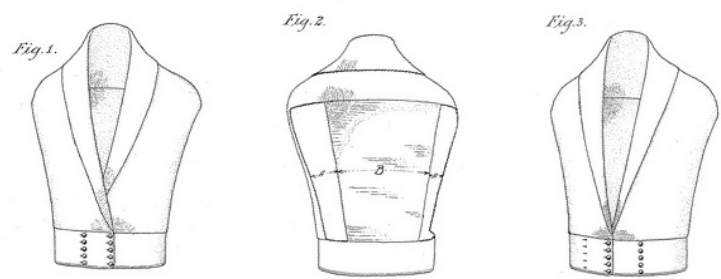

*Figure 7: Einstein's design of a blouse.*

**4. Conclusions**

Einstein's inventions and patents are mostly of historical importance now. His inventions were over-shadowed by technological advancement in 20[th] century. However, that Einstein attempted and successfully designed those implements; automatic camera, refrigerator, pump, sound system, even a blouse, serve to show how diversified and talented was the great scientist. History behind those inventions showed that he was not just a theoretical physicist, sitting on an ivory tower, rather his foot was firmly rooted on the earth. He was troubled by human miseries and tried his best to relieve them.

We may mention that in recent years, interest in Einstein-Szilard refrigerator has been revived. In 20th century, Chlorofluorocarbons (CFCs) were used widely as refrigerant, but they damage the environment by depleting ozone layers. With increased awareness of environment and associated climate change, efforts are on for Green technology. CFCs are being replaced by hydrofluorocarbons (HFCs) as they are less damaging to the environment. Malcolm McCulloch, an electrical engineer at Oxford, is trying to bring Einstein's refrigerator back as it is environmentally friendly.  McCulloch believed that efficiency of the design can be improved by tweaking of the gases. He is also trying to use solar energy for the heat source required in the refrigerator. Einstein-Szilard refrigerator has certain advantage that make them particularly useful in developing countries, where demand for cooling appliances is quickly increasing. The appliance requires minimum maintenance as no moving part is involved. Furthermore, many areas of under-developed countries are still without electricity and the refrigerator can met their cooling requirement, as it is not dependent on electricity. The refrigerator will be extremely beneficial to the vaccination programme in under-developed countries, where vast areas are still without electricity. Vaccines are required to be stored at 2-8 degree centigrade temperature. Vaccination programme in those areas suffer for want of appropriate storage facility. Einstein-Szilard refrigerator can be used to store the vaccines and change the quality of life.